\documentclass[twocolumn,amsmath,amssymb,prl,superscriptaddress]{revtex4-1}

\usepackage{graphicx}
\usepackage{dcolumn}
\usepackage{bm}

\begin{document}

\title{Jamming, relaxation and crystallization of a super-cooled fluid \\ 
in a three-dimensional lattice}
\author{H. Levit}
\affiliation{Beverly and Raymond Sackler School of Physics and Astronomy,
Tel Aviv University, Tel Aviv 69978, Israel}
\author{Z. Rotman}
\affiliation{Beverly and Raymond Sackler School of Physics and Astronomy,
Tel Aviv University, Tel Aviv 69978, Israel}
\affiliation{Department of Cell Biology and Physiology, Washington University School of Medicine, St. Louis, Missouri 63110}
\author{E. Eisenberg}
\affiliation{Beverly and Raymond Sackler School of Physics and Astronomy,
Tel Aviv University, Tel Aviv 69978, Israel}

\begin{abstract}
Off-equilibrium dynamics of a three-dimensional lattice model with nearest- and next nearest-neighbors exclusions is studied. At equilibrium, the model undergoes a first-order fluid-solid transition. Non-equilibrium filling, through random sequential adsorption with diffusion, creates amorphous structures and terminates at a disordered state with random closest packing density that lies in the equilibrium solid regime. The approach towards random closest packing is characterized by two distinct power-law regimes, reflecting the formation of small densely packed grains in the long time regime of the filling process. We then study the fixed-density relaxation of these amorphous structures towards the solid phase. The route to crystallization is shown to deviate from the simple grain growth proposed by classical nucleation theory. Our measurements suggest that relaxation is driven mainly by coalescence of neighboring crystallized grains which exist in the initial amorphous state.
\end{abstract}


\maketitle
\section*{Introduction}

Random sequential adsorption (RSA) processes are well-studied examples of off-equilibrium dynamics \cite{evans93}. In these processes, particles are sequentially introduced into a system at random locations, and get adsorbed if this location is compatible with the other particles. For models with inter-particle hard-core exclusion, the process ends when all deposition options are exhausted. Typically, RSA processes terminate at a limiting density far lower than $\rho_{CP}$, the closest packing density of the respective system. For example, random deposition of dimers on a one-dimensional (1D) lattice creates sites whose two nearest neighbors are filled. These sites can never be filled and the process jams with terminal density $\theta_J=0.864664$ \cite{flory39}.

When adsorbed particles are allowed to diffuse, jammed states can be relaxed by particle movement (RSAD models). Jamming density in RSAD models is naturally higher and in some cases closest packing density is reached. In some cases, such as the 1D lattice RSAD and in the square lattice exclusion models N1 and N2 (particles deposited on a square lattice, where each particle excludes deposition on its four nearest-neighbors (N1) or up to second nearest neighbors (N2)), the density converges to $\rho_{CP}$ via a $t^{-1/2}$ power law  \cite{wnp93m,wnp93p,eisbar98}. In contrast, in the 2D N3 lattice model (exclusion up to third nearest neighbors) the density converges via a  $t^{-1}$ power law to a limiting density lower than $\rho_{CP}$, which is termed the {\it random closest packing density} $\rho_{RCP}=0.171626< \rho_{CP}=0.2$ \cite{eisbar99}.

Random closest packing states are characterized by the absence of mobile particles (but a few rattlers). Thus, diffusive relaxation from these off-equilibrium states is impossible. However, off-equilibrium diffusive relaxation can be studied by terminating RSAD filling process at a predetermined density, lower than the random closest packing one, and following relaxation dynamics towards crystallization. Better characterization of the crystallization process is of interest (see e.g. \cite{binsta76,heerklei83,Auerfren01}), as some of its aspects, most notably the crystallization rate, are not well explained by standard nucleation theory \cite{Auerfren01}. In addition, recent works have highlighted the possibility that while classical nucleation theory describes well the nucleation at concentrations slightly above the coexistence region, highly saturated super-cooled fluids undergo different mechanisms of
nucleation. These include spinodal nucleation and a third mechanism, yet to be fully understood, which is characterized by small displacements of the particles \cite{pusey1, pusey3}.

\begin{figure}[h!]
\includegraphics[width=4cm,height=4cm,angle=0]{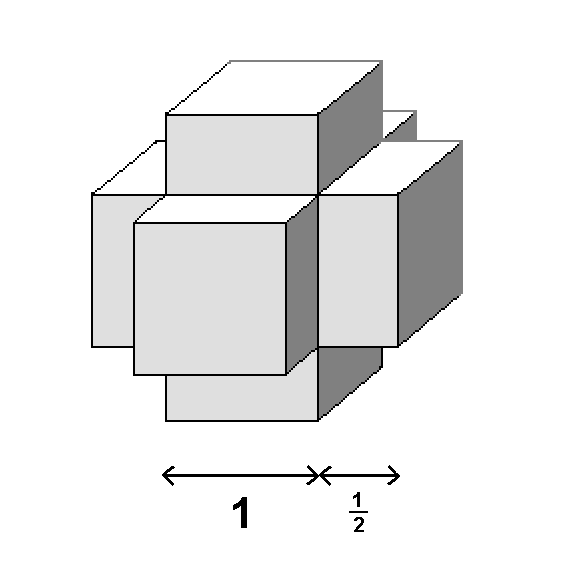}
\includegraphics[width=4cm,height=4cm,angle=0]{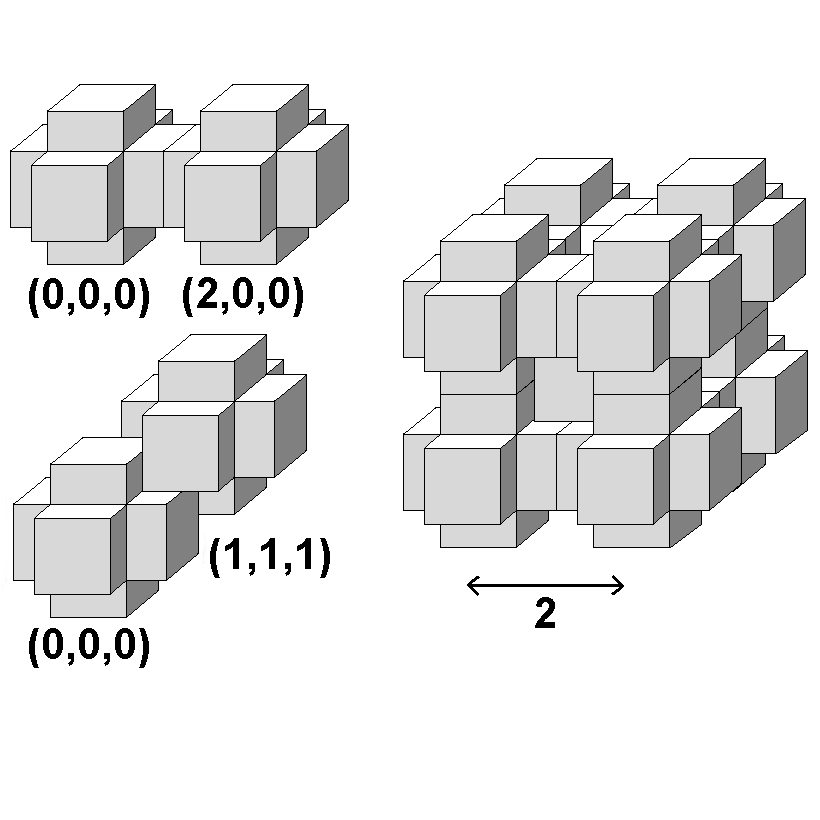}
\caption{A hard core particle which enforces the model's exclusion rules (left). The central cube is one cubic-lattice site and the whole particle spans 2 lattice sites in each direction. Several configurations are shown in order to demonstrate the allowed nearest neighbors and the closest packing configuration - a BCC lattice (right).}
\label{model000}
\end{figure}

Here we study a three-dimensional (3D) model on the cubic lattice, where the inter-particle interaction is limited to  hard-core exclusion up to the second-nearest neighbors, henceforth termed the 3DN2 model. This can be thought of as a system of hard particles of volume $4$ sites, whose closest packing configuration is a body-centered cubic (BCC) arrangement with a lattice constant equal to 2 lattice sites (Fig. \ref{model000}). The closest packing {\it particle density} is therefore $\rho_{CP}=0.25$. In the following we measure all densities relative to this maximal particle density ($n=\rho /\rho_{CP}$), i.e. we use {\it volume density}. We start by exploring the equilibrium behavior of the 3DN2 model, and show that it undergoes a first-order phase transition, with a density gap between $n_f=0.413$ and $n_s=0.519$. We then look at the RSAD dynamics, and show that unlike the 2D N3 model, where the random closest packing density lies within the density gap, the RSAD dynamics of 3DN2 terminates at density $n_{RCP}=0.81$, well within the solid phase. Thus, RSAD dynamics enables us to compress the system deep into the super-cooled regime, while keeping the system microscopically amorphous. 

Finally, we study the diffusive relaxation of the amorphous state as a function of density. 
The study of the off-equilibrium relaxation of disordered super-cooled configurations towards the equilibrium solid has attracted much interest, mainly in the context of super-cooled glass-formers near the glass transition \cite{ediger,cavagna,berthier,roteis09}. In contrast, here this relaxation process leads into the equilibrium crystal. Yet, it turns out that several characteristics of the dynamics are reminiscent of glassy relaxation. In fact, it has been argued that in a number of systems an ideal glass transition could be related to the kinetic spinodal \cite{pusey1,2dlj,swift,cavspin1, cavagna03,roteis10}, in the sense that spontaneous crystallization of many stable micro-crystallites could lead to a kinetic arrest due to the competition between the different crystalline orders. In comparison, the 3DN2 system presents no such competition, and thus the kinetic spinodal is manifested only in the form of a modified solidification process. Indeed, we find that the route to crystallization in the high densities, well above the equilibrium transition density, is characterized by most particles moving as little as needed to shift to the sublattice on which the system crystallizes, in agreement with recent works on glassy hard spheres system \cite{pusey3}. Furthermore, we observe that a leading mechanism is the alignment of neighboring micro-crystallites, which exist in the amorphous {\it initial state}, against each other.

\begin{figure}
\includegraphics[width=7cm,height=5cm,angle=0]{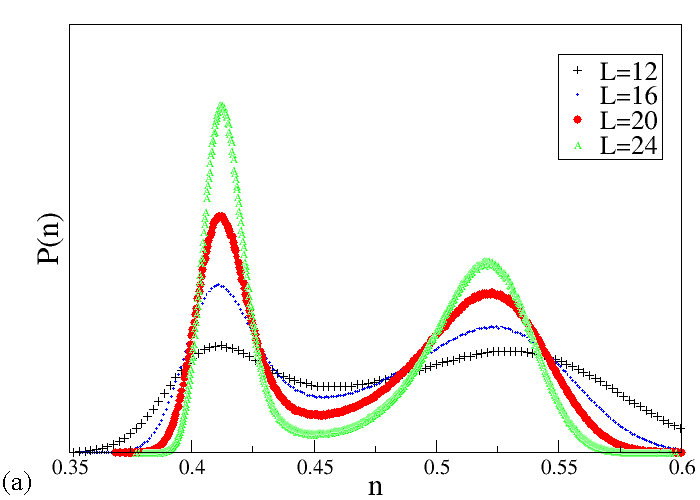}\\
\includegraphics[width=7cm,height=5cm,angle=0]{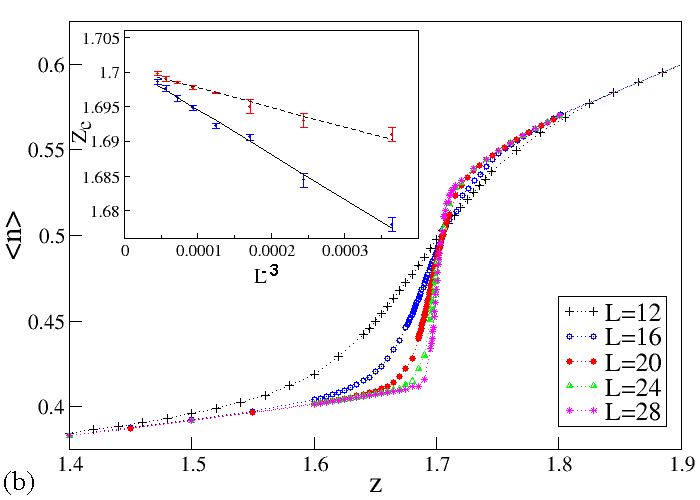}\\
\includegraphics[width=7cm,height=5cm,angle=0]{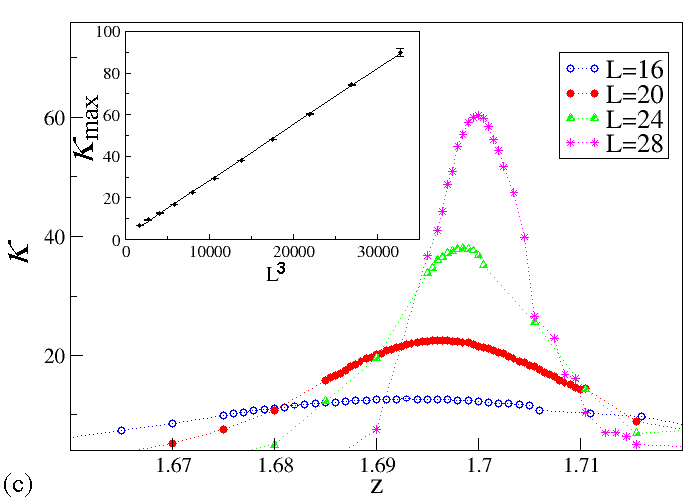}\\
\includegraphics[width=7cm,height=5cm,angle=0]{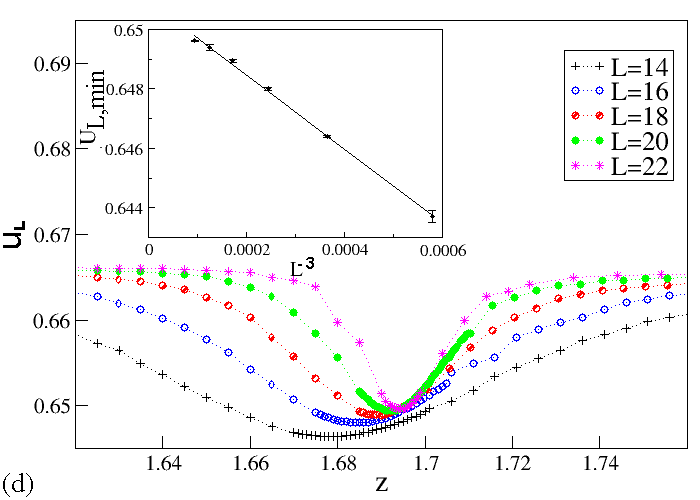}\\
\caption{(Color online) The phase transition. (a) A histogram of the density at constant $z\simeq z_{c}$ showing the system divides its time between two, high (solid) density and low (liquid) density, states, meaning the transition is of first order. (b) The average density in the vicinity of the transition as measured in lattices of different linear size. The inset shows the finite size scaling of $z_{c}$ according to the locations of the peaks in the compressibility (dashed line) and of the dips in Binder's cumulant (Eq. \ref{binc}) (solid line). (c) and (d) are the measured $\kappa(z)$ and $U_L(z)$, respectively, and (in the insets) the finite size scaling of their critical values.}
\label{thermo_pt}
\end{figure}
\section*{Thermodynamics}

To characterize the model's thermodynamics, we used Monte-Carlo (MC) simulations of a grand canonical ensemble. In each MC step, particles are allowed to be adsorbed on the lattice (while adhering to the exclusion rules) or desorbed from it. The probability of adsorbing/desorbing a particle is governed by the chemical potential $\mu$. We started the simulations with an empty lattice and raised the activity $z=e^{\beta\mu}$ monotonically (compressing, or cooling), equilibrating the system after each raise. At the critical regime, the distribution of the density for a finite system (at a constant $z$) is bimodal (Figure \ref{thermo_pt}(a)), reflecting the transitions between two phases of different density - higher density solid and lower density fluid. Thus, the transition is a first order one. 

We used finite size scaling, with lattices of sizes $L=12,14,16...$ $30$, to extrapolate the transition parameters at the thermodynamic limit \cite{binlan84}. This analysis (Figure \ref{thermo_pt}) verified that the transition is indeed first order.
The compressibility, 
\begin{equation}
\kappa =\left( \langle n^2\rangle -\langle n\rangle ^2 \right) \cdot L^3,
\end{equation}
exhibits a rounded peak at the transition, whose height is proportional to the volume, as expected for a first-order transition. Binder's cumulant, the forth order moment of the form 
\begin{equation}
U_L=1-\frac{\langle n^4\rangle }{3\cdot \langle n^2\rangle ^2},
\label{binc}
\end{equation}
also conforms to the first-order behavior - it is constant far from the transition, and near it it has a dip whose height is proportional to $L^{-3}$ \cite{binder81}. The location of the transition, can be extracted from either the positions of $\kappa$'s maxima or of $U_L$'s minima. Both converge (the latter faster than the first) towards a value of $z_c=1.7010(5)$. The limiting densities of the liquid and solid phases are those around which the distribution of the density peaks, $n_f=0.4131\pm0.003$ and $n_s=0.5195\pm0.0005$ accordingly, in agreement with a previous thermodynamic study \cite{Panagiotopoulos}. Here too, the finite size correction exhibits the non-anomalous $O(L^{-3})$ behavior.

\section*{Approach to Random Closest Packing}

We have studied the RSAD dynamics \cite{eisbar98} of the 3DN2 system in extensive MC simulations, using lattices of linear size between $64$ and $256$ (over $3$ million particles). RSAD dynamics is equivalent to a grand canonical dynamics with an infinite chemical potential, resulting in particle deposition whenever possible, and no desorption.  As the only energy scale in the problem is the chemical potential, infinite chemical potential is equivalent to an infinitely fast cooling. First, we performed random sequential deposition on the (initially empty) lattice. This filling process happens at time zero and terminates at density $0.49362\pm0.00004$, in the equilibrium coexistence regime. When all adsorption possibilities are exhausted, the system is let to diffuse, i.e. particles attempt to move to a randomly selected nearest neighbor site. Whenever diffusion creates a space large enough for a particle to be added without violating the exclusion rules, a particle is immediately placed there. A rejection-free algorithm was used to reduce run-time and allow simulation of larger lattices. 

\begin{figure}
\includegraphics[width=8cm,height=5.3cm,angle=0]{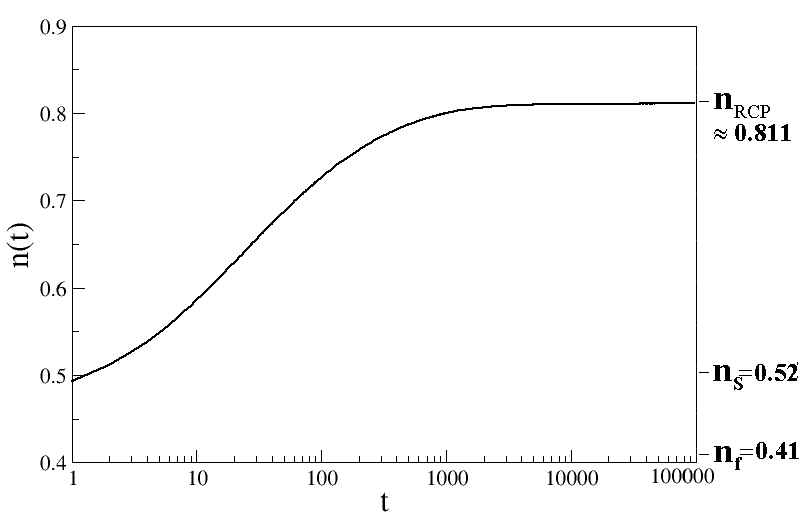}
\caption{Random sequential adsorption with diffusion (RSAD). The volume density is plotted vs. time on a logarithmic scale. The results are averaged over 1500 realizations, for lattice of size $L=256$ (finite size effects were minute, thus we present only the results for $L=256$). Relative standard error is less than $0.0005$ for all $t$.}
\label{rcp256}
\end{figure}

\begin{figure}
\includegraphics[width=8cm,height=5.3cm]{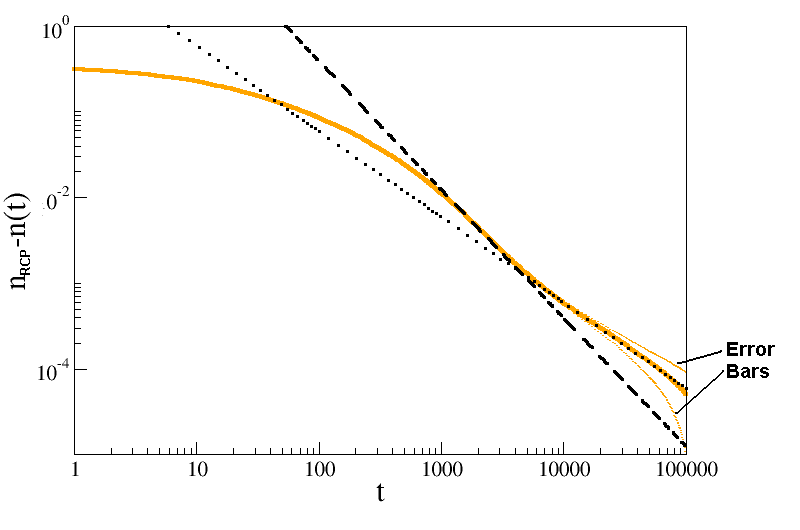}
\caption{(Color online) Density vs. time along the RSAD process. The residual density (distance, in density, to the termination random closest packing density) is plotted, in order to highlight the two power-law regimes: the earlier with a residual density proportional to $t^{-3/2}$ (dashed line), and the later one exhibiting a $t^{-1}$ behavior (dotted line).}
\label{rcp256fit}
\end{figure}

Measurements of volume density (Fig. \ref{rcp256}) clearly show that the process jams at a limiting density $n_{rcp}= 0.81095\pm0.00004$, lower than the closest packing value, but much higher than the transition region. The long-time approach towards $n_{rcp}$ follows the asymptotics $n(t)=n_{rcp}-A/t$ (see Fig. \ref{rcp256fit}). However, at shorter times, one observes a regime in which the residual vacancies density follows a $t^{-3/2}$ rule. The long-time RSAD dynamics can be described by a model of holes diffusion (facilitated by diffusion of the neighboring particles) \cite{eisbar97}. The vacancies in the lattice can be relaxed by particle deposition only when multiple holes colocalize, which can be described as an annihilation process from the holes' perspective. The simplified diffusion-annihilation description predicts that the vacancies density should follow a $t^{-3/2}$ law in 3D ($t^{-1}$ in 2D), according to the probability of two diffusing holes to meet (or the return probability of a random walk). Here we find that the intermediate-time behavior agrees with hole diffusion in 3D while the long-time $1/t$ behavior suggests that hole diffusion is effectively 2D. Analyzing the microscopic processes, we found that this effective dimensional reduction results from the formation of stable, perfectly packed, clusters. The holes left in the systems are trapped between clusters and perform 2D diffusive motion along the boundaries of the clusters, resulting in a slower density gain, $n_{rcp}-n(t)\sim t^{-1}$, associated with the effectively 2D diffusion. A similar phenomenon was observed for the 2D N1 and N2 models, where long-time RSAD relaxation was governed by 1D processes \cite{eisbar98}.

\begin{figure}
\includegraphics[width=8cm,height=5.3cm]{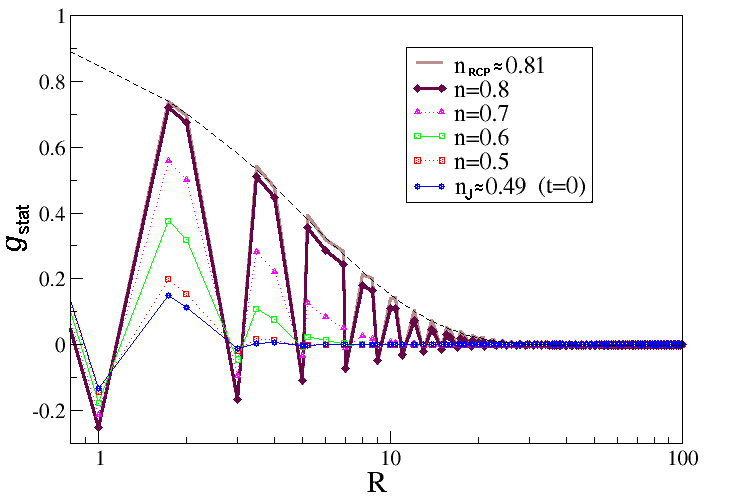}
\caption{(Color online) The static correlation function $g_{stat}({\bf R})$ at different times along the filling process, taken along the principal axes and main diagonals and plotted against $R$=$|{\bf R}|$. The positive peaks (at $R$'s fitting the closest packing configuration) decay exponentially with distance. The correlation length reaches $\lambda \simeq 5.3$ sites at $\rho_{RCP}$ (regardless of lattice size), as shown by the fit to a pure exponent (dashed line). 
}
\label{rcpgs}
\end{figure}

The existence of packed clusters at long times is corroborated by measurements of the static density-density correlation function \begin{equation}
g_{stat}({\bf R})=\frac{\langle \rho({\bf r})\rho({\bf r+R})\rangle - \langle \rho({\bf r}) \rangle^2}{\rho(1-\rho)},
\end{equation} 
where 
\begin{equation}
\rho=\frac{1}{V}\sum_{{\bf r}_i} \rho({\bf r}_i,t)
\end{equation}
is the total (time independent) number density.
Taken along the principal axes and main diagonals, the correlation function is characterized by the peaks and dips manifesting the closest packing BCC order in short range: peaks for even integers and dips for odd integers (principal axes) as well as peaks for multiples of $\sqrt{3}$ (diagonals). The closely-packed clusters grow in the short time regime until they are stopped by neighboring clusters residing in other sublattices. The decay of the correlation function shows that the typical size of the clusters towards the end of the filling process is about $5$ sites. The clusters reach their maximum size when the density is a little over $0.8$, coinciding with the later time regime when density gain starts converging through the $t^{-1}$ rule (Fig. \ref{rcpgs}). 

\section*{Relaxation Dynamics}

\begin{figure}
\includegraphics[width=7.5cm,height=5.3cm]{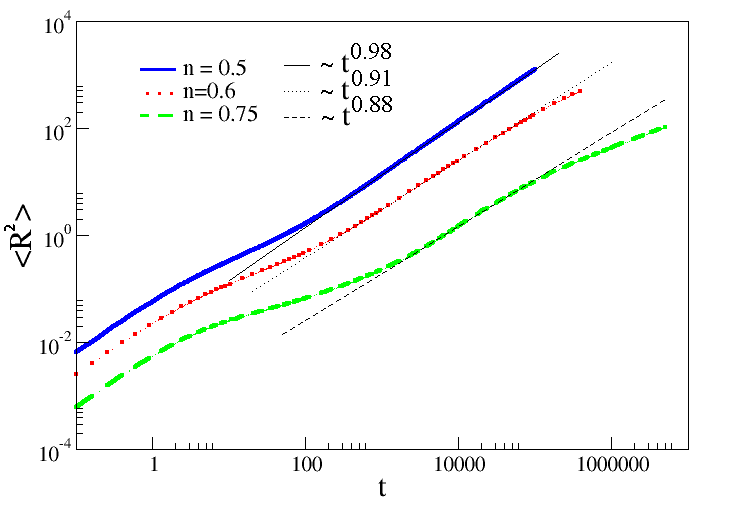}
\caption{(Color online) The mean squared displacement averaged over all particles in ten realizations of a system of size $L=128$, as a function of density. The log-log plot highlights the diffusive regimes: in times of order of a single time step, the particles perform free diffusion within cages. As the cages' sizes are of order of $1$ site, this is followed by a sub diffusive stage, where the particles are still mostly bound to their cages, and vibrate around their origins. At later times, the diffusivity is dominated by the movement of particles between cages and greatly depends on density. For $10^3<t<10^5$, we see that at densities within the transition gap the behavior is again diffusive (with a diffusion constant ~100 times smaller than that of the first regime), but systems that were super-cooled into the solid regime exhibit sub-diffusive behavior, characterized by a power-law $<R^2>=Dt^{\alpha}$ with $\alpha<1$. Finally, for long times, diffusion is diminished as the systems goes through solidification (The longer times are inaccessible in the gap densities as some of the particles already reach the limit of the finite system.}
\label{dynR2128}
\end{figure}

As the random closest packing density lies deep within the solid phase, RSAD dynamics may be used to create a disordered configurations at densities deep within the equilibrium solid regime. In order to study off-equilibrium relaxation of the super-cooled fluid, we therefore employ RSAD dynamics to fill the system up to some fixed density, and then let the system relax diffusionally, keeping the number of particles fixed. 
An important characterization of the dynamics is given by particles' mobility. Off-equilibrium systems are often characterized by a sub-diffusive behavior. For example, some glassy systems exhibit caged dynamics where particles are trapped in small regions, within which diffusion takes place. Such systems display a short-time diffusion within the cages, followed by a period characterized by sub-diffusion as a result of the localization of the particles. The transition rates between cages determines the long-range, long-time, mobility and could be either diffusive or sub-diffusive depending on the nature of the barriers between the cages. Here, the average squared displacement displays a similar behavior, as shown in Fig. \ref{dynR2128}, only that the later stages are inter-mixed with the solidification process itself. The system starts at a disordered state so that the particles are caged and localized - free brownian movement is only measured at very short times, of order of one time unit, meaning the average cage size is of order a single lattice site. Then, after some (density-dependent) time, structural relaxation begins with movement between cages, and the diffusion measured is an anomalous diffusion resulting from the distribution of cages' size.
The higher the density, the lower is the exponent characterizing the sub-diffusive motion (see  Fig. \ref{dynR2128}). Finally, as the system crystallizes, particles get localized permanently and mobility diminishes.

\begin{figure}
\includegraphics[width=7.5cm,height=5.3cm]{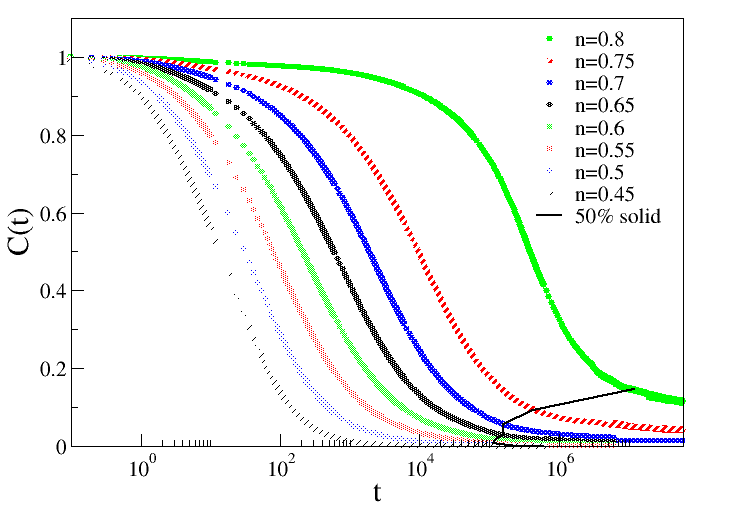}
\caption{(Color online) $C(t)$ in an $L=128$ Lattice, for densities ranging from $0.45$ to $0.8$. The results are an average over several hundred realizations (there are less realizations in the long time regime, hence the larger error bars). Black line indicating the time at which half of the particles are on the same sublattice.}
\label{ctc128}
\end{figure}

The evolution of the relaxation process can be quantitatively studies through the density-density autocorrelation function

\begin{equation}
C(t)=\frac{1}{\rho(1-\rho)}\left( \frac{1}{V}\sum_{{\bf r}_i} 
\rho({\bf r}_i,0)\rho({\bf r}_i,t)-\rho^2\right),
\end{equation}
where ${\bf r}_i$ runs over all $V$ sites, and $\rho({\bf r}_i,t)$ is one (zero) if at time t there is 
(not) a particle at site ${\bf r}_i$. This correlation function (or rather its long-time limit) is often used as the Edwards-Anderson order parameter in glassy materials. Beyond the glass transition, finite correlations are expected even for infinitely long times.
Apparently, the model exhibits the two-step relaxation that is typical of glassy systems (Fig. \ref{ctc128}). Due to the aforementioned caging phenomenon, there is very little relaxation in the short time regime as the particles mostly move back and forth within their cages. This leads to the formation of a plateau in the autocorrelation function, known as the $\beta$-regime. As density increases, the plateau extends to longer times. It is followed by the structural relaxation in the $\alpha$-regime, where cages are broken and restructured, and the correlation function decays to zero. While some of the simulations ended having finite $C(t)$ a detailed finite size scaling shows that this is a finite size effect. The $\alpha$-decay takes the form of a stretched exponential, expressing the spatially heterogeneous character of the system.

Despite the apparent similarities to the behavior of glassy systems, it should be stressed that the long-time relaxation observed here is qualitatively different from the alpha-phase relaxation seen in glassy materials. As indicated in Fig. \ref{ctc128}, the long time relaxation overlaps with the solidification of the system. The nature of this process and the reason it is accompanied by spatially long-ranged relaxation can be understood looking at the microscopic picture.

\begin{figure}
\includegraphics[width=7.5cm,height=5.3cm]{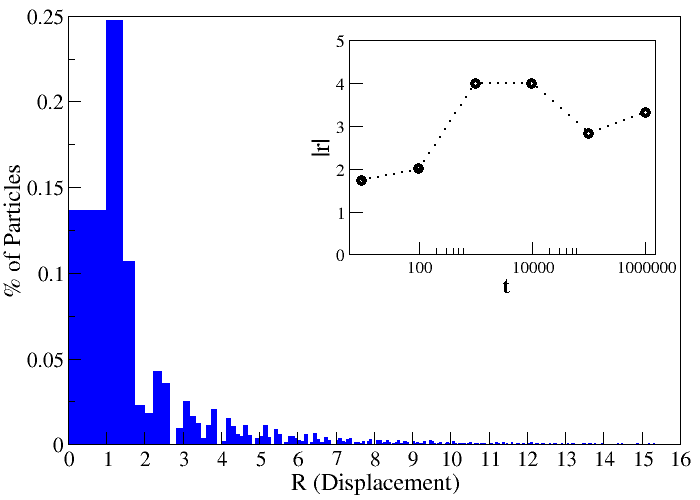}
\caption{(Color online) Particles' displacements in a high-density ($n=0.75$) system at a time it has already crystallized (almost all particles are lying on the same BCC sublattice). Most particles have moved less than $\sqrt{3}$ sites.The inset shows the displacement correlation length as estimated by the decay of the displacement correlation function to $0.2$. Displacement correlation is short ranged even for times longer than crystallization time. Data presented for a system of size $L=128$ and density $n=0.75$ (over 100 realizations), and is typical for other high densities and reasonable correlation cutoffs.}
\label{dyndisplace}
\end{figure}


\section*{Microscopic Picture of the Crystallization Dynamics}

Detailed analysis of microscopic processes underlying the dynamics reveals some interesting features of the relaxation process. We focus on the total displacements of particles along the process (Fig. \ref{dyndisplace}), and find that in the high-density regime a considerable number of the particles move only a distance of $1$ to $\sqrt{3}$ sites away from their $t=0$ location, i.e. up to one shift along each of the axes. Moreover, the particles' movement distance is highly correlated between neighboring particles, i.e. different regions in the model have their own characteristic distance. The sizes of these clusters can be quantified by the spatial correlations of the displacement 
\begin{equation}
C_D({\bf r},t)=\langle R({0,t})R({\bf r},t) \rangle - \langle R \rangle^2, 
\end{equation}
which exhibit short-range correlations. In order to obtain an estimate for the length scale characterizing these displacement correlation we look at the distance at which $C_D({\bf r},t)$ decays to an arbitraty cutoff value ($0.2$). The inset of Fig.\ref{dyndisplace} presents the time-dependence of this length scale. Not only are the estimated correlation lengths short-ranged compared to the static correlation measured in t=0, but they also do not grow over time. It is important to notice that even though the exact value of the length scale is as arbitrary as the threshold used, these two observations are not.

The above observations are inconsistent with simple nucleation theory picture of crystallization, according to which growing nucleation seeds are continuously fed by their fluid environment. Instead, we observe the alignment of micro-crystallites (present at $t=0$) to be the main mechanism leading to crystallization. The particles forming the boundaries of the clusters are effectively caged by the surrounding clusters, so that their movement is somewhat restricted (resulting in the sub-diffusive behavior described above). This microscopic picture fits with the global dynamic measurements described above: the displacement correlation length observed above reflect nothing but the size of the clusters moving cooperatively to re-align themselves against each other. 
The crystallization mechanism described here, characterized by minute displacements of particles along the process, is reminiscent of a similar phenomenon recently observed for the hard spheres glassy system \cite{pusey3}. It is believed that the transition to this regime has to do with the density being so high (or the super-cooled fluid being over-saturated) such that the kinetic spinodal is reached. Under these conditions, 
micro-crystallites of any size are stable. Thus, virtually all of the micro-crystallites naturally occurring in the initial amorphous state survive the relaxation process, and crystallization is achieved by re-alignment of these against each other. This highlights another possible connection to the glassy behavior. It was suggested that in some systems an ideal glass transition occurs at the kinetic spinodal, as micro-crystallites having competing crystal order are all stable and thus no global order may be achieved \cite{2dlj,swift,cavspin1,cavagna03,roteis10}. The fate of a deeply super-cooled fluid is then determined by the nature of the micro-crystallites occurring when kinetic spinodal is hit. If they are easily alignable, one would end up having a solid, while when they are not easily alignable, a stable amorphous glass would emerge.

\begin{figure}
\includegraphics[width=7cm,height=5cm,angle=0]{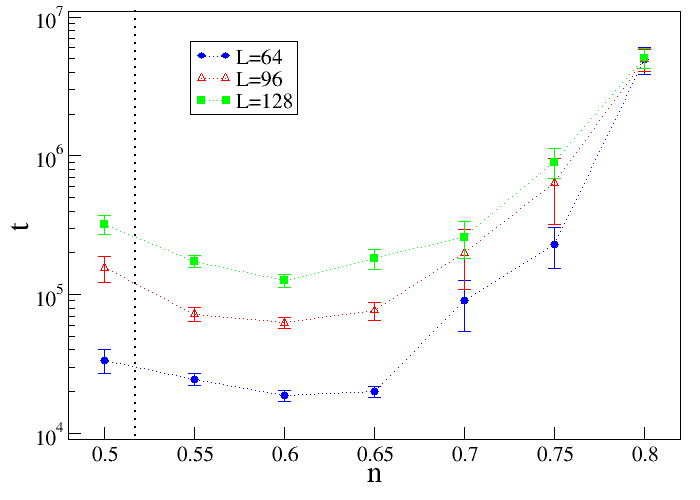}
\caption{(Color online) The time of solidification, defined as the first time at which more than half of the particles are on the same sublattice, averaged over $77$ realizations, as a function of system size and density. Realizations in which two or more macroscopic domains are formed are averaged only if one of the domains contains more than half of all particles.}
\label{dynsolidt}
\end{figure}

Finally, we comment about the crystallization time. We define the crystallization time as the time (measured from the end of the RSAD process) till half of the particles occupy the same sublattice (The equilibrium occupation fraction of the leading sublattice depends only weakly on the density, changing from $0.9$ at the onset of crystallinity to $0.96$ for $n=0.6$ and approaching unity for high densities). Measuring this time as a function of density, one observe a non-monotonous picture (Fig. \ref{dynsolidt}). As density increases, the preference of the solid phase over the disordered phases increases. At the same time, the dynamics becomes increasingly sluggish. The result of these two effects combined is the non-monotonous behavior. The long relaxation time in high-density systems also stems from the fact that they are prone to the formation of two or more macroscopic domains which require a long time to align with one another. The increase in the average solidification time is enhanced by the finite size effects - the larger the lattice, the longer it takes for the system to solidify. 

\section{Conclusion}

The 3DN2 model presented here exhibits a non trivial RCP dynamics. The density gain in the long-time regime, controlled by holes' diffusion, is going through two distinguished regimes corresponding to an unordered dense liquid state, followed by a state of  many micro-crystallites in which the particles mainly perform 2D diffusion on the clusters' boundaries. Keeping the density fixed, the system relaxes to a solid. In the high-density regime, beyond the kinetic spinodal, even small micro-crystallites are stable. Relaxation is then brought about through alignment of the small solid clusters against each other. It then follows that the dynamics of the system is heterogeneous, similarly to super-cooled glass-formers. The crystallization process, is dramatically different than the one described by classical nucleation theory.


\end{document}